\documentclass[aps,showpacs,amsfonts,nofootinbib,preprintnumbers,nobalancelastpage]{revtex4}

\usepackage[english]{babel}
\usepackage{amsmath,amssymb}
\usepackage[dvips]{graphicx}
\usepackage[sort&compress]{natbib}
\usepackage{bbm}
\usepackage{color}
\usepackage{ulem}

\DeclareMathAlphabet{\mathsc}{OT1}{cmr}{m}{sc}

\allowdisplaybreaks
\def\321{SU(3) $\otimes$ SU(2) $\otimes$ U(1)}

\def\lsim{\raise0.3ex\hbox{$\;<$\kern-0.75em\raise-1.1ex\hbox{$\sim\;$}}}
\def\gsim{\raise0.3ex\hbox{$\;>$\kern-0.75em\raise-1.1ex\hbox{$\sim\;$}}}

\newcommand{\flux}[2][]{\ensuremath{\ifthenelse{\equal{#1}{}}{}{^{#1}\!}\mathit{#2}}}

\newcommand{\BR}{\mathrm{BR}}

\newcommand{\AddrAHEP}{%
  AHEP Group, Instituto de F\'{\i}sica Corpuscular --
  C.S.I.C./Universitat de Val{\`e}ncia \\
  Edificio Institutos de Paterna, Apt 22085, E--46071 Valencia, Spain}

\begin{document}

%------include the correct preprint number
\preprint{IFIC/12-45}

\title{Probing neutralino properties in minimal supergravity\\
  with bilinear $R$-parity violation}

\author{F.\ de Campos}
\email{camposc@feg.unesp.br}
\affiliation{Departamento de F\'{\i}sica e Qu\'{\i}mica,
             Universidade Estadual Paulista, Guaratinguet\'a -- SP,  Brazil }

\author{O.\ J.\ P.\ \'Eboli}
\email{eboli@fma.if.usp.br}
\affiliation{Instituto de F\'{\i}sica,
             Universidade de S\~ao Paulo, S\~ao Paulo -- SP, Brazil \\
            Institut de Physique Th\'{e}orique, CEA-Saclay
Orme des Merisiers, 91191 Gif-sur-Yvette, France
}

\author{M.\ B.\ Magro}
\email{magro@fma.if.usp.br}
\affiliation{Instituto de F\'{\i}sica,
             Universidade de S\~ao Paulo, S\~ao Paulo -- SP, Brazil.}
\affiliation{Centro Universit\'ario Funda\c{c}\~ao Santo Andr\'e,
             Santo Andr\'e -- SP, Brazil.}

\author{W.\ Porod} 
\email{porod@physik.uni-wuerzburg.de}
\affiliation{Institut f\"ur Theoretische Physik und
             Astronomie, Universit\"at W\"urzburg, Germany}

\author{D.\ Restrepo}
\email{restrepo@uv.es}
\affiliation{Instituto de F\'{\i}sica, Universidad de Antioquia - Colombia}

\author{S.\ P.\ Das}
\email{spdas@ific.uv.es}
\affiliation{\AddrAHEP}

\author{M.\ Hirsch}
\email{hirsch@ific.uv.es}
\affiliation{\AddrAHEP}

\author{J.\ W.\ F.\ Valle}
\email{valle@ific.uv.es}
\affiliation{\AddrAHEP}

%---------------------------------------------------------------------
\begin{abstract}

  Supersymmetric models with bilinear $R$-parity violation (BRPV)
  can account for the observed neutrino masses and mixing parameters
  indicated by neutrino oscillation data.  We consider minimal
  supergravity versions of BRPV where the lightest supersymmetric
  particle (LSP) is a neutralino. This is unstable, with a large
  enough decay length to be detected at the CERN Large Hadron Collider
  (LHC).
  We analyze the LHC potential to determine the LSP properties, such
  as mass, lifetime and branching ratios, and discuss their relation
  to neutrino properties.

\end{abstract}
%---------------------------------------------------------------------


\maketitle

%%%%%%%%%%%%%%%%%%%%%%%%%%%%%%%%%%%%%%%%%%%%%%%%%%%%%%%%%%%%%%%%%%%%%%
\section{Introduction}

Elucidating the electroweak breaking sector of the Standard Model (SM)
constitutes a major challenge for the Large Hadron Collider (LHC) at
CERN. Supersymmetry provides an elegant way to stabilize the Higgs
boson scalar mass against quantum corrections provided supersymmetric
states are not too heavy, with some of them expected within reach for
the LHC.  Searches for supersymmetric particles constitute a major
item in the LHC agenda~\cite{Chatrchyan:2011zy,daCosta:2011qk,%
ATLAS:2011ad,Aad:2011qa,Aad:2011cwa,Aad:2011zj,Khachatryan:2011tk,%
Chatrchyan:2011bz,Chatrchyan:2011wba,Chatrchyan:2011ff}, as
many expect signs of supersymmetry (SUSY) to be just around the
corner.  However the first searches up to $\sim$ 5 fb$^{-1}$ at the
LHC interpreted within specific frameworks, such as Constrained
Minimal Supersymmetric Standard Model (CMSSM) or minimal supergravity
(mSUGRA) indicate that squark and gluino masses are in excess of $\sim
1$ TeV
\cite{Bechtle:2012zk}. \smallskip

Despite intense efforts over more than thirty years, little is known
from first principles about how exactly to realize or break
supersymmetry.  As a result one should keep an open mind as to which
theoretical framework is realized in nature, if any.  Supersymmetry
search strategies must be correspondingly re-designed if, for example,
supersymmetry is realized in the absence of a conserved R
parity \cite{ATLAS:2011ad,Aad:2011zb}.\smallskip

Another major drawback of the Standard Model is its failure to account
for neutrino oscillations~\cite{nakamura2010review,Maltoni:2004ei},
whose discovery constitutes one of the major advances in particle
physics of the last decade.
An important observation is that, if supersymmetry is realized without
a conserved R parity, the origin of neutrino masses and mixing may be
intrinsically
supersymmetric~\cite{aulakh:1982yn,hall:1983id,ross:1984yg,Ellis:1984gi}.

Indeed an attractive dynamical way to generate neutrino mass at the
weak scale is through non-zero vacuum expectation values of \321
singlet scalar
neutrinos~\cite{masiero:1990uj,romao:1992vu,romao:1997xf}. This leads
to the minimal effective description of R parity violation, namely
BRPV~\cite{hirsch:2004he}. In contrast to the simplest variants of the
seesaw mechanism~\cite{valle:2006vb} such supersymmetric alternative
has the merit of being testable in collider experiments, like the
LHC~\cite{decampos:2005ri,decampos:2007bn,decampos:2008re,DeCampos:2010yu}.
Here we analyze the LHC potential to determine the lightest neutralino
properties such as mass, decay length and branching ratios, and
discuss their relation to neutrino properties.
\smallskip

%%%%%%%%%%%%%%%%%%%%%%%%%%%%%%%%%%%%%%%%%%%%%%%%%%%%%%%%%%%%%%%%%%%%%%
\section{Bilinear $\mathbf{R}$-parity violating SUSY models}
\label{sec:model}

The bilinear R-Parity violating models are characterized by two
properties: first the usual MSSM R-conserving superpotential is
enlarged according to~\cite{diaz:1997xc}
\begin{equation}
  W_{\text{BRPV}} = W_{\text{MSSM}}  + \varepsilon_{ab}
\epsilon_i \widehat L_i^a\widehat H_u^b \; ,
\label{eq:sp:BRPV}
\end{equation}
where there are 3 new superpotential parameters ($\epsilon_i$), one
for each fermion generation~\footnote{In a way similar to the $\mu$
  term in the MSSM superpotential, the required smallness of the
  bilinear parameters $\epsilon_i$ could arise dynamically, through a
  nonzero vev, as in \cite{masiero:1990uj,romao:1992vu,romao:1997xf,nilles:1996ij}
  and/or be generated radiatively~\cite{giudice:1988yz}.}.  The second
modification is the addition of an extra soft term
\begin{equation}
V_{\text{soft}} = V_{\text{MSSM}} - \varepsilon_{ab} 
B_i\epsilon_i\widetilde L_i^aH_u^b
\end{equation}
that depends on three soft mass parameters $B_i$. For the sake of
simplicity we considered the R-conserving soft terms as in minimal
supergravity (mSUGRA). Notice that the presence of the new soft
interactions prevents the new bilinear terms in Eq.~(\ref{eq:sp:BRPV})
to be rotated away \cite{diaz:1997xc}. \smallskip

The new bilinear terms break explicitly R parity as well as lepton
number and induce non-zero vacuum expectation values $v_i$ for the
sneutrinos. As a result, neutrinos and neutralinos mix at tree level
giving rise to one tree--level neutrino mass scale, which we identify
with the atmospheric scale. The other two neutrino masses are generated
through loop diagrams~\cite{Hirsch:2000ef,diaz:2003as}. This model
provides a good description of the observed neutrino oscillation
data~\cite{Maltoni:2004ei}. \smallskip

The BRPV--mSUGRA model is defined by eleven parameters
\begin{equation}
m_0\,,\, m_{1/2}\,,\, \tan\beta\,,\, {\mathrm{sign}}(\mu)\,,\,
A_0 \,,\,
\epsilon_i \: {\mathrm{, and}}\,\, B_i\,,
\end{equation}
where $m_{1/2}$ and $m_0$ are the common gaugino mass and scalar soft
SUSY breaking masses at the unification scale, $A_0$ is the common
trilinear term, and $\tan\beta$ is the ratio between the Higgs field
vacuum expectation values (vevs). In our analyzes the new parameters
($\epsilon_i$ and $B_i$) are determined by the neutrino masses and
mixings, therefore, we have only to vary the usual mSUGRA
parameters. For the sake of simplicity in what follows we fix $A_0 =
-100$ GeV, $\tan\beta = 10$ and ${\mathrm{sign}}(\mu)> 0$ and present
our results in the plane $m_0 \otimes m_{1/2}$.  \smallskip

\begin{figure}
\includegraphics[width=0.49\textwidth,angle=0]{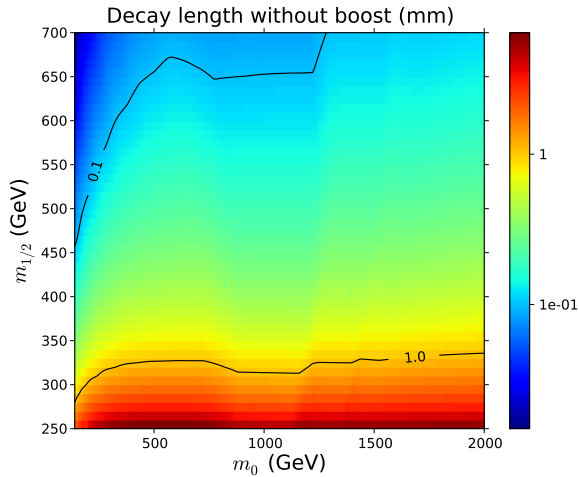}
\caption{Lightest neutralino decay length as a function of mSUGRA
  parameters $m_0$ and $m_{1/2}$, for $A_0 = -100$ GeV, $\tan\beta =
  10$  and ${\mathrm{sign}}(\mu)> 0$. }
\label{fig:decaylength}
\end{figure}

Due to the smallness of the neutrino masses, the BRPV interactions
turn out to be rather feeble, consequently the LSP has a lifetime long
enough that its decay appears as a displaced vertex. We show in
figure~\ref{fig:decaylength} the LSP decay length as a function of
$m_0$ and $m_{1/2}$, when the remaining values for sign$(\mu),~A$ and
$\tan\beta$ are taken as mentioned above. Therefore, we can anticipate
that the LSP decay vertex can be observed at the LHC within a large
fraction of the parameter space. \smallskip

%%%%%%%%%%%%%%
%%%%%%%%%%%%%%
 
Depending on the SUSY spectrum the lightest neutralino decay
channels include fully leptonic decays
\[
\tilde{\chi}^0_1 \to \nu \ell^+ \ell^-   
 \;\;\;\;,\;\;\;\;
\tilde{\chi}^0_1 \to \nu \tau^+ \tau^-
 \;\;\;\;\hbox{ and }\;\;\;\;
\tilde{\chi}^0_1 \to \nu \tau^\pm \ell^\mp \,
\]
with $\ell =e$  or $\mu$; as well as semi-leptonic decay modes 
\[
\tilde{\chi}^0_1 \to \nu q \bar{q}
 \;\;\;\;,\;\;\;\;
\tilde{\chi}^0_1 \to \tau q^\prime \bar{q}
 \;\;\;\;,\;\;\;\;
\tilde{\chi}^0_1 \to \ell q^\prime \bar{q}
 \;\;\;\;\hbox{ and }\;\;\;\;
\tilde{\chi}^0_1 \to \nu b \bar{b} \;.
\]
If kinematically allowed, some of these modes take place via two--body
decays, like $\tilde{\chi}^0_1 \to W^\mp \mu^\pm$, $\tilde{\chi}^0_1
\to W^\mp \tau^\pm$, $\tilde{\chi}^0_1 \to Z \nu$, or
$\tilde{\chi}^0_1 \to h \nu$, followed by the $Z$, $W^\pm$ or $h$
decay; for further details see Ref.~\cite{porod:2000hv,decampos:2007bn}.  In
addition to these channels there is also the possibility of the
neutralino decaying invisibly into three neutrinos, however, this
channel  reaches at most a few per-cent
\cite{porod:2000hv}\footnote{However,
in models where a Majoron is present, it can be dominant 
\cite{GonzalezGarcia:1991ap,Bartl:1996cg,hirsch:2006di,hirsch:2008ur}}. \smallskip

\begin{figure}
\includegraphics[width=0.48\textwidth]{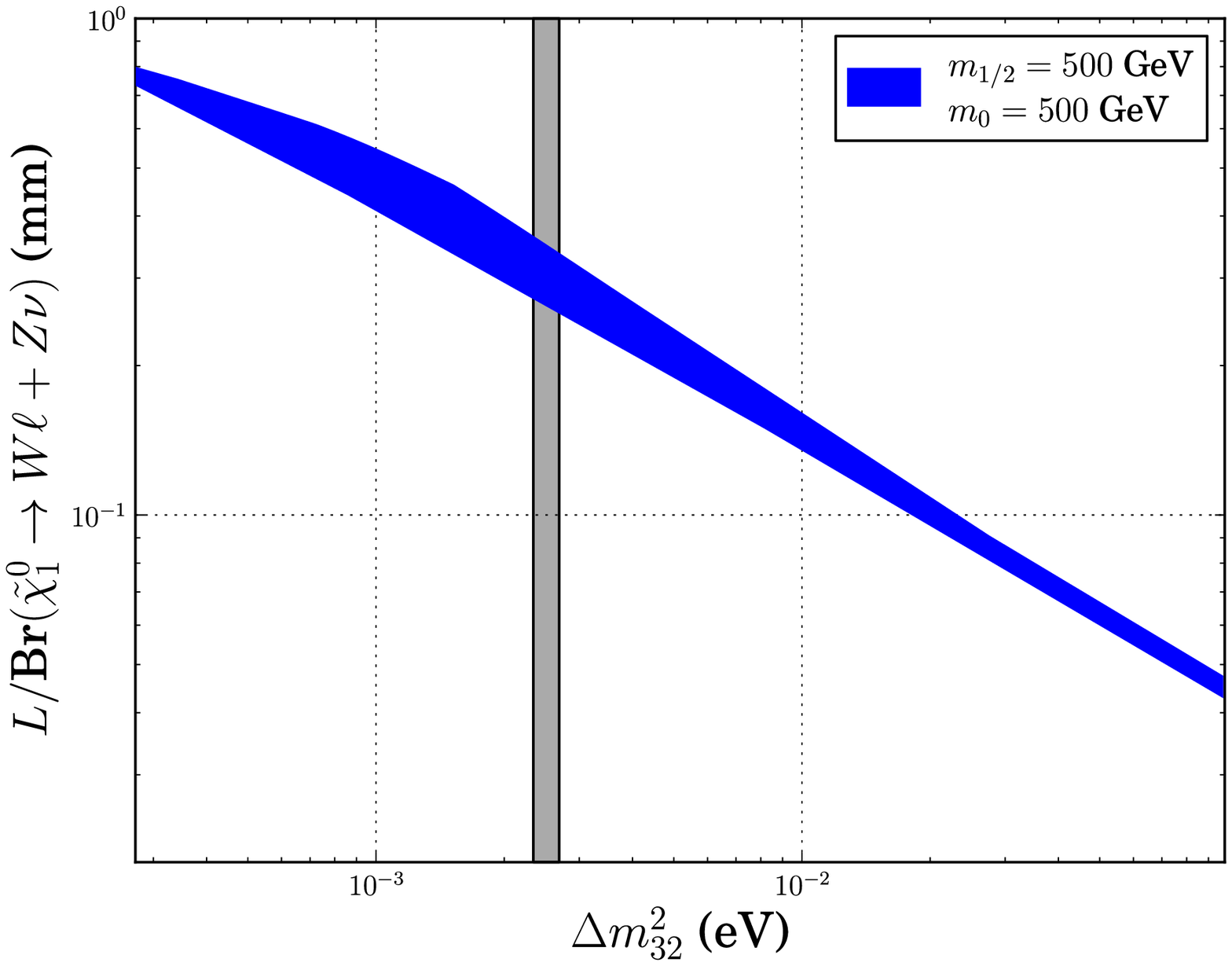}
%LoverWlZnu_Dm23_500_500.eps}
\hfill
\includegraphics[width=0.48\textwidth]{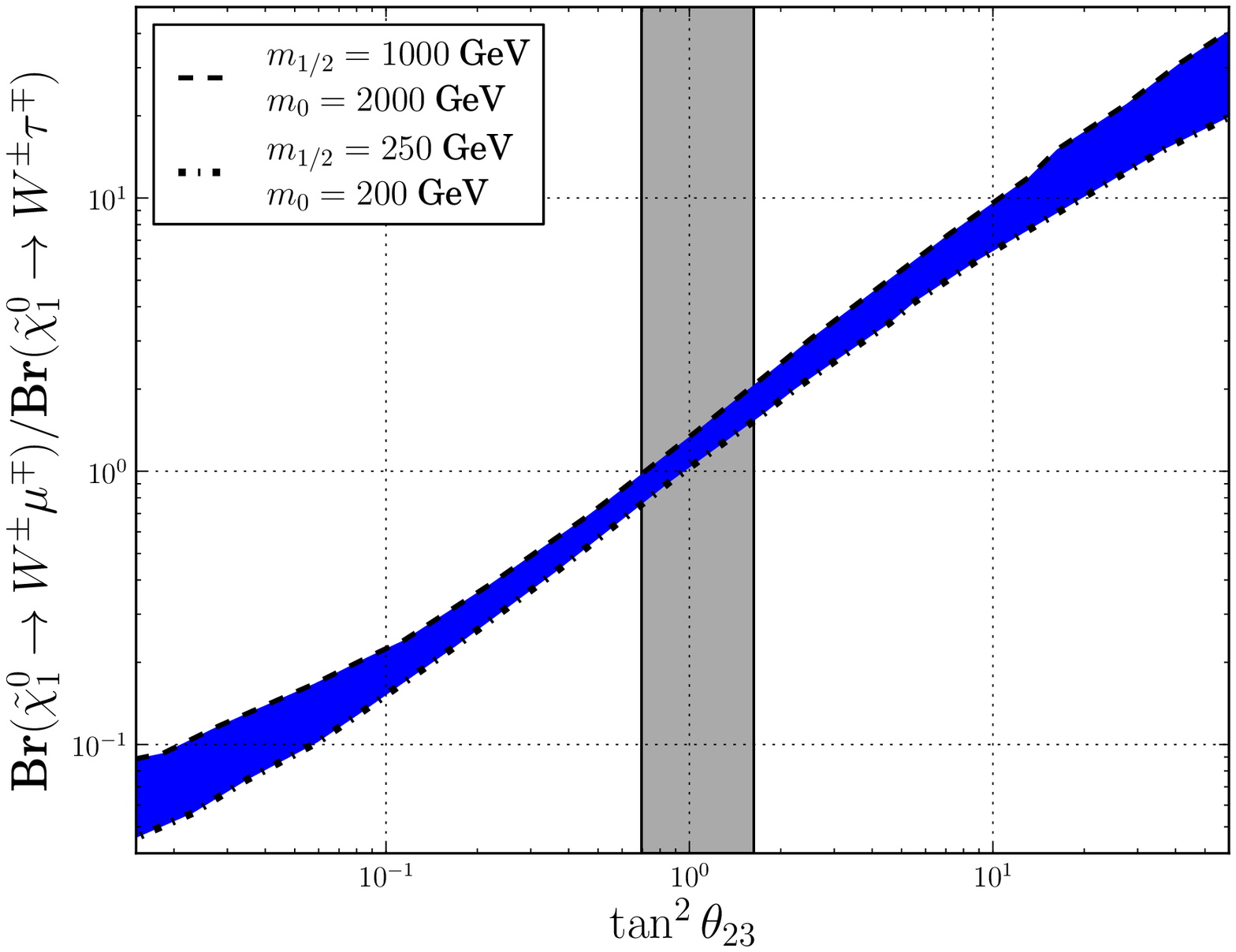}
\caption{Correlating LSP decay properties with neutrino oscillation
  parameters.  The left panel shows the connection between the
  displayed LSP decay length parameter and the atmospheric squared
  mass scale $\Delta m^2_{32}$. The right panel depicts the relation
  between $\hbox{Br}(\tilde{\chi}^0_1 \to W^\pm \mu^\mp)/
  \hbox{Br}(\tilde{\chi}^0_1 \to W^\pm \tau^\mp) $ and the atmospheric
  mixing angle.  The vertical shaded bands indicate the 2$\sigma$
  allowed values of the corresponding neutrino oscillation
  parameters~\cite{Tortola:2012te}. }
\label{fig:correlations}
\end{figure}

Neutrino masses and mixings as well as LSP decay properties are
determined by the same interactions, therefore, there are connections
between high energy LSP physics at the LHC and neutrino oscillation
physics. For instance, the ratio between charged current decays
\begin{equation}
   \frac{ \hbox{Br} ( \tilde{\chi}^0_1 \to W^\pm \mu^\mp)}
{ \hbox{Br} ( \tilde{\chi}^0_1 \to W^\pm \tau^\mp)}
\end{equation}
is directly related to the atmospheric mixing angle
\cite{datta:1999xq}, as illustrated in
the right panel of figure~\ref{fig:correlations}; this relation was
already considered in Ref.~\cite{DeCampos:2010yu}.  The vertical bands
in figure~\ref{fig:correlations} correspond to the latest 2$\sigma$ precision in the
determination of $\theta_{23}$ and $\Delta m_{32}^2$ from
Ref.~\cite{Tortola:2012te}.  \smallskip

Another interesting interconnection between LSP properties and
neutrino properties is the direct relation between neutrino mass
squared difference $\Delta m^2_{32}$ and the ratio
\begin{equation}
      R_{32} = \frac{L_0}{\hbox{Br} ( \tilde{\chi}^0_1 \to W \ell 
                    + Z\nu)}
\label{eq:ratio}
\end{equation}
 as is illustrated in left panel
of figure~\ref{fig:correlations}. 
 Here $L_0$ is the LSP decay length and one has to sum over
all leptons and neutrinos in the final states.
 One can understand this relation in the following way. In the BRPV 
model the tree-level neutrino mass is proportional to 
$m_{\nu}^{\rm Tree} \propto |\Lambda|^2$, where $|\Lambda|^2 = 
\sum_i\Lambda_i^2$, with $\Lambda_i = \epsilon v_d + \mu v_i$, 
is the so-called alignment vector. Couplings between the 
gauginos and gauge bosons plus leptons/neutrinos are proportional 
to  $\Lambda_i$ as well \cite{porod:2000hv}. Thus, one expects 
that after summing over the lepton generations the partial 
width of the neutralino into gauge bosons is also proportional 
to $|\Lambda|^2$. The decay length is the inverse width and 
dividing by the branching ratio into gauge boson final states 
picks out the partial width of the neutralino into gauge bosons. 
This leads to the correlation of $R_{32}$ with the atmospheric 
neutrino mass scale, since $m_{ \rm Atm}$ is identified mostly 
with $m_{\nu}^{\rm Tree}$, apart from some minor 1-loop corrections.   
  \smallskip
 
%%%%%%%%%%%%%%
%%%%%%%%%%%%%%

%%%%%%%%%%%%%%%%%%%%%%%%%%%%%%%%%%%%%%%%%%%%%%%%%%%%%%%%%%%%%%%%%%%%%%
\section{Analyses framework and basic cuts}
\label{ana:frame}

Our analyzes aim to study the LHC potential to probe the LSP
properties exploring its detached vertex signature. We simulated the
SUSY particle production using PYTHIA version
6.408~\cite{Sjostrand:2000wi,Sjostrand:1993yb} where all the
properties of our BRPV-mSUGRA model were included using the SLHA
format~\cite{Skands:2003cj}. The relevant masses, mixings, branching
ratios, and decay lengths were generated using the SPHENO
code~\cite{Porod:2003um,Porod:2011nf}. \smallskip

In our studies we used a toy calorimeter roughly inspired by the
actual LHC detectors. 
We assumed that the calorimeter coverage is $|
\eta | < 5$ and that its segmentation is $\Delta \eta \otimes \Delta
\varphi = 0.10 \times 0.098$. The calorimeter resolution was included
by smearing the jet energies with an error
\[
   \frac{\Delta E}{E} = \frac{0.50}{\sqrt{E}} \oplus 0.03 \; .
\]
Jets were reconstructed using the cone algorithm in the subroutine
PYCELL with $\Delta R = 0.4$ and jet seed with a minimum transverse
energy $E_{T,min}^{cell} = 2$ GeV. \smallskip

Our analyzes start by selecting events that pass some typical triggers
employed by the ATLAS/CMS collaborations, {\em i.e.}  an event to be
accepted should fulfill  at least one of the following requirements:
\begin{itemize}

\item the event contains one electron or photon with $p_T > 20$ GeV;

\item the event has an isolated muon with $p_T > 6$   GeV;

\item the event exhibits two isolated electrons or photons with $p_T > 15$ GeV;

\item the event has one jet with transverse momentum in excess of 100 GeV;

\item the events possesses missing transverse energy greater than 100 GeV.

\end{itemize}

We then require the existence of, at least, one displaced vertex that
is more than $5\sigma$ away from the primary
vertex~\cite{decampos:2007bn} -- that is, the detached vertex is
outside the ellipsoid
\begin{equation}
  \label{eq:minellip}
      \left ( \frac{x}{5\delta_{xy}} \right )^2
   +  \left ( \frac{y}{5\delta_{xy}} \right )^2
   +  \left ( \frac{z}{5\delta_{z}} \right )^2   = 1 \; ,
\end{equation}
where the $z$-axis is along the beam direction. We used the ATLAS
expected resolutions in the transverse plane ($\delta_{xy} = 20~\mu$m)
and in the beam direction ($\delta_z = 500~\mu$m). To ensure a good
reconstruction of the displaced vertex we further required that the
LSP decays within the tracking system {\em i.e.}  within a radius of
$550$ mm and $z$--axis length of $3000$ mm. In our model the decay
lengths are such that this last requirement is almost automatically
satisfied; see figure~\ref{fig:decaylength}. \smallskip

%%%%%%%%%%%%%%%%%%%%%%%%%%%%%%%%%%%%%%%%%%%%%%%%%%%%%%%%%%%%%%%%%%%%%%
\section{LSP mass measurement}
\label{sec:lsp-mass-measurement}

In order to accurately measure the LSP mass from its decay products we
focused our attention on events where the LSP decays into a charged
lepton ($e^\pm$ or $\mu^\pm$) and a $W$ that subsequently decays into
a pair of jets. In addition to the basic cuts described above we
further required charged leptons to have
\begin{equation}
  p_T^\ell > 20 \hbox{ GeV and }  |\eta_\ell| < 2.5 \; .
\end{equation}
We demanded the charged lepton to be isolated, {\em i.e.} the sum of
the transverse energy of the particles in a cone $\Delta R = 0.3$
around the lepton direction should satisfy
\begin{equation}
  \sum_{\Delta R < 0.3} E_T < 5 \hbox{ GeV} \; .
\end{equation}
We identified the hadronically decaying $W$ requiring that its decay
jets are central 
\begin{equation}
   p_T^j > 20 \hbox{ GeV} \;\;\;\;,\;\;\;\;
   | \eta_j| < 2.5 \; ,
\end{equation}
and  that their invariant mass is compatible with the
$W$ mass:
\begin{equation}
   | \eta_j| < 2.5  \;\;\;\; \hbox{and} \;\;\;\;
   | M_{jj} - M_W | < 20 \hbox{ GeV} \; .
\end{equation}

\begin{figure}
\includegraphics[width=0.49\textwidth]{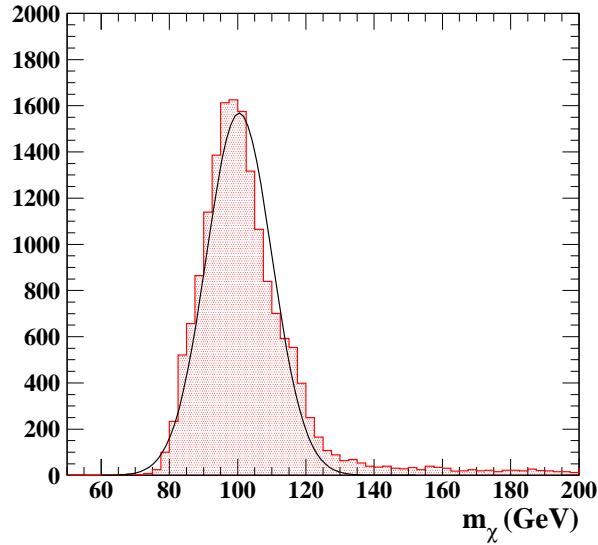}
\caption{Illustration of the lepton--jet--jet invariant mass spectrum
  fitted to obtain the LSP mass.  In this figure we considered $m_0 =
  250$ GeV, $m_{1/2} = 250$ GeV, $\tan \beta = 10$, $A_0 = -100$ GeV,
  and $\hbox{sgn}(\mu) > 0$ which leads to a LSP mass of 101 GeV.}
\label{fig:lspmass}
\end{figure}

In order to obtain the LSP mass, we considered points in the $m_0
\otimes m_{1/2}$ plane with more than 10 expected events for an
integrated luminosity of 100 fb$^{-1}$. We have performed a Gaussian
fit to the lepton--jet--jet invariant mass; as an illustration of the
lepton--jet--jet invariant mass spectrum see figure~\ref{fig:lspmass}.
As we can see from this figure, the actual LSP mass (101 GeV) is with
1\% of its fitted value (100.4 GeV).  \smallskip

%%%%%%%%%%%%%%
%%%%%%%%%%%%%%

\begin{figure}
\includegraphics[width=0.48\textwidth]{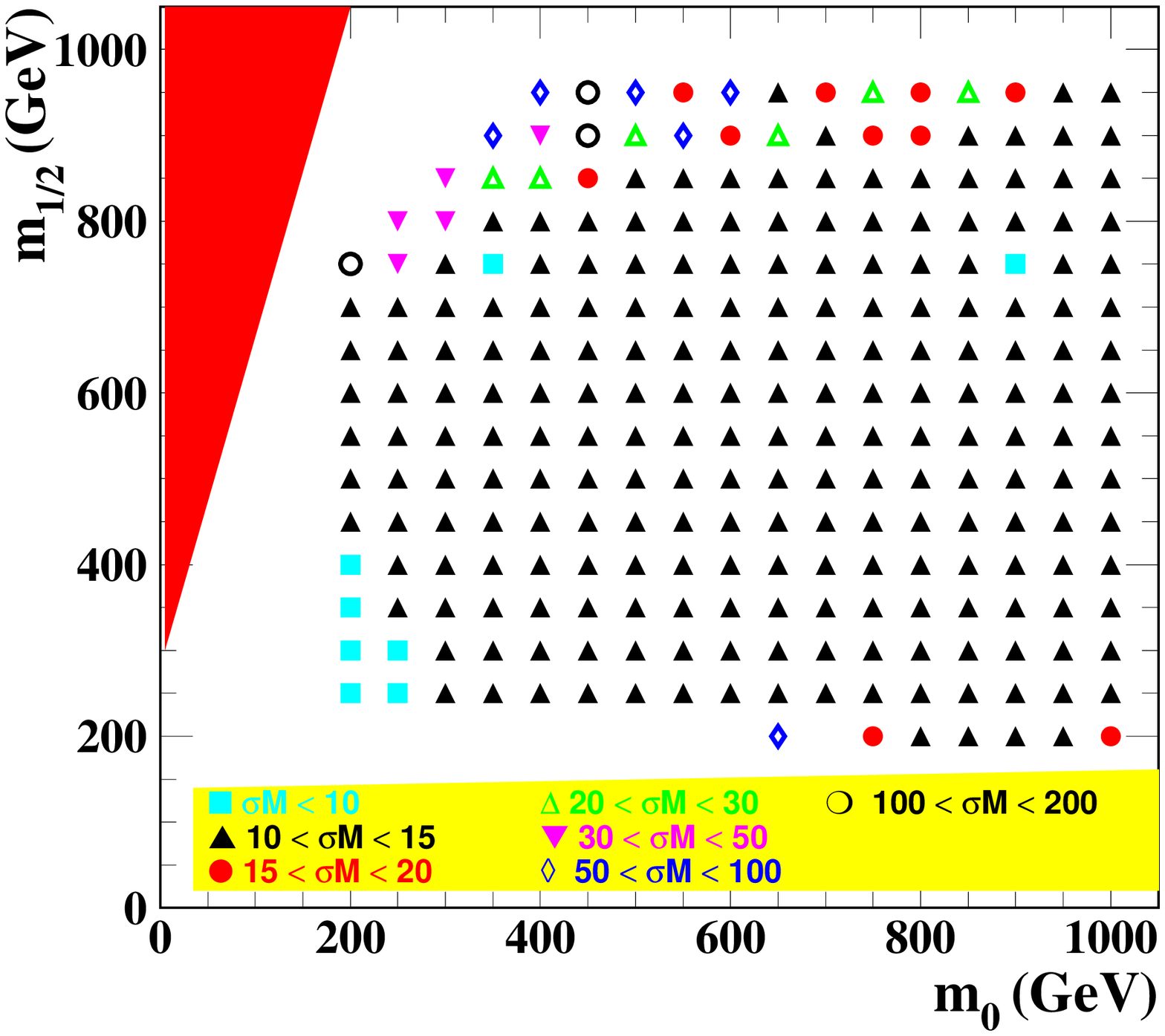}
\hfill
\includegraphics[width=0.48\textwidth]{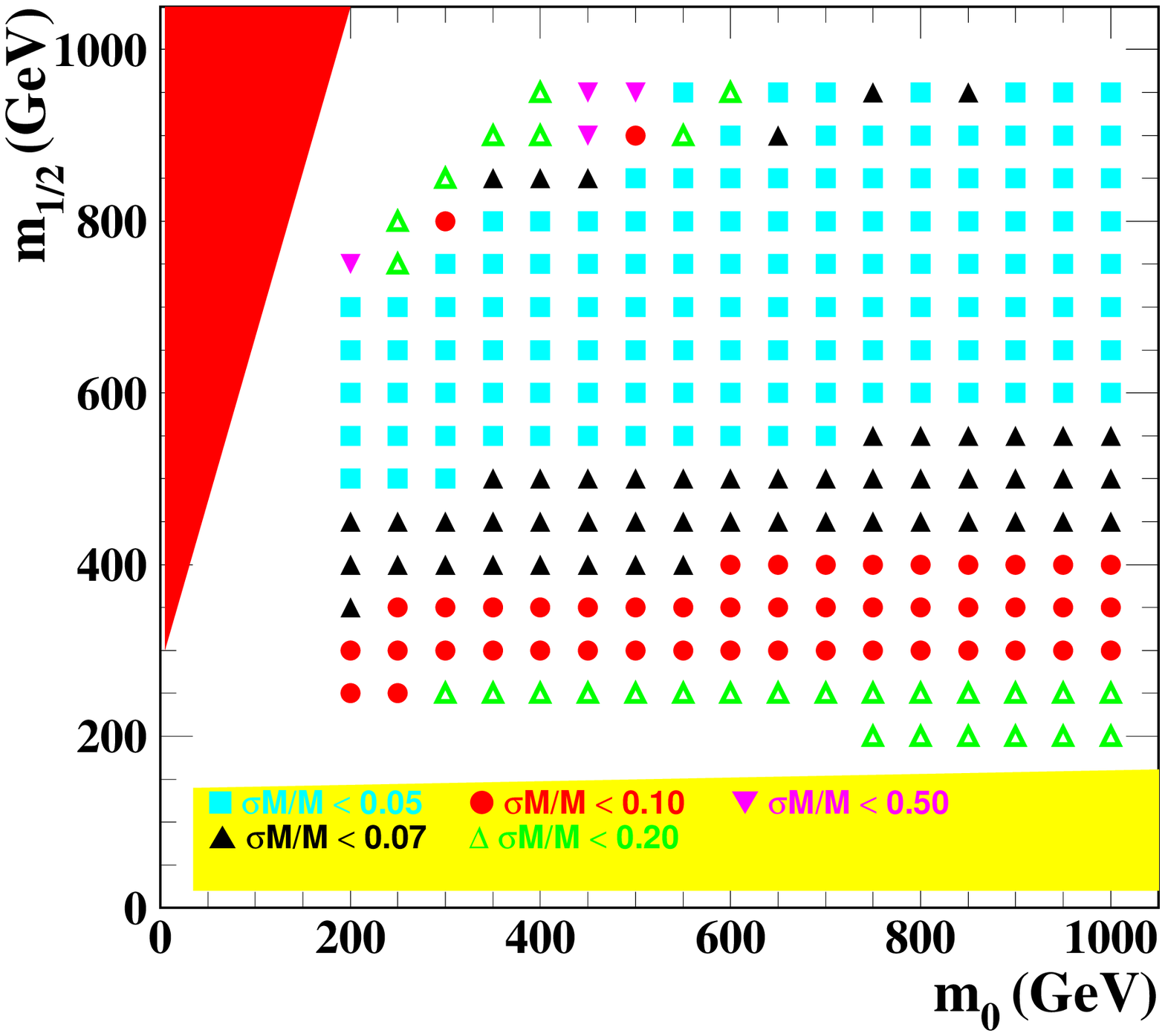}
\caption{ The left panel presents the error ($\sigma_M$) on the LSP
  mass as a function of the $m_0 \otimes m_{1/2}$ point for $A_0 =
  -100$ GeV, $\tan \beta = 10$, $\hbox{sgn}\mu > 0$ and an integrated
  luminosity of 100 fb$^{-1}$, while the right panel displays the
  relative error in the LSP mass determination
  $\sigma_M/M_{\tilde{\chi}^0_1 }$.  }
\label{fig:mlsp_errors}
\end{figure}

In order to better appreciate the precision with which the LSP mass
can be determined for other choices of mSUGRA parameters we have
repeated the analysis for a wide grid of values in the $m_0 \otimes
m_{1/2}$ plane.  The left panel of figure~\ref{fig:mlsp_errors}
depicts the achievable precision in the LSP mass measurement for an
integrated luminosity of 100 fb$^{-1}$ as a function of $m_0 \otimes
m_{1/2}$ for $A_0 = -100$ GeV, $\tan \beta = 10$, and $\hbox{sgn}(\mu)
>0$. As one can see the LSP mass can be measured with an error between
10 and 15 GeV within a sizeable fraction of the ($m_0 \otimes
m_{1/2}$) plane. Only at high $m_{1/2}$ there is a degradation of the
precision due to poor statistics.  The right panel in
figure~\ref{fig:mlsp_errors} shows that indeed this is enough to
determine the LSP mass to within 5 to 10\% in a relatively wide chunk
of parameter space. \smallskip

%%%%%%%%%%%%%%

%%%%%%%%%%%%%%

%%%%%%%%%%%%%%%%%%%%%%%%%%%%%%%%%%%%%%%%%%%%%%%%%%%%%%%%%%%%%%%%%%%%%%
\section{ LSP decay length measurement}
\label{sec:lsp-decay-length}

Another important feature of the LSP in our BRPV-mSUGRA model is its
decay length (lifetime). Within the simplest mSUGRA bilinear R parity
violating scheme this is directly related to the squared mass
splitting $\Delta m_{32}^2$, well measured in neutrino oscillation
experiments~\cite{Tortola:2012te}. In this analysis we
  considered events where the LSP decay contains
 at least three
  charged tracks, {\em i.e.}  the LSP decays into $\ell jj$,
  with $\ell =e$ or $\mu$. Here we sum over all jets
  as well as over $\tilde \chi^0_1 \to \ell W \to \ell jj$
   and all 
  three body decays leading to the same final state.\smallskip

%%%%%%%%%%%%%%%
\begin{figure}
\includegraphics[width=0.48\textwidth]{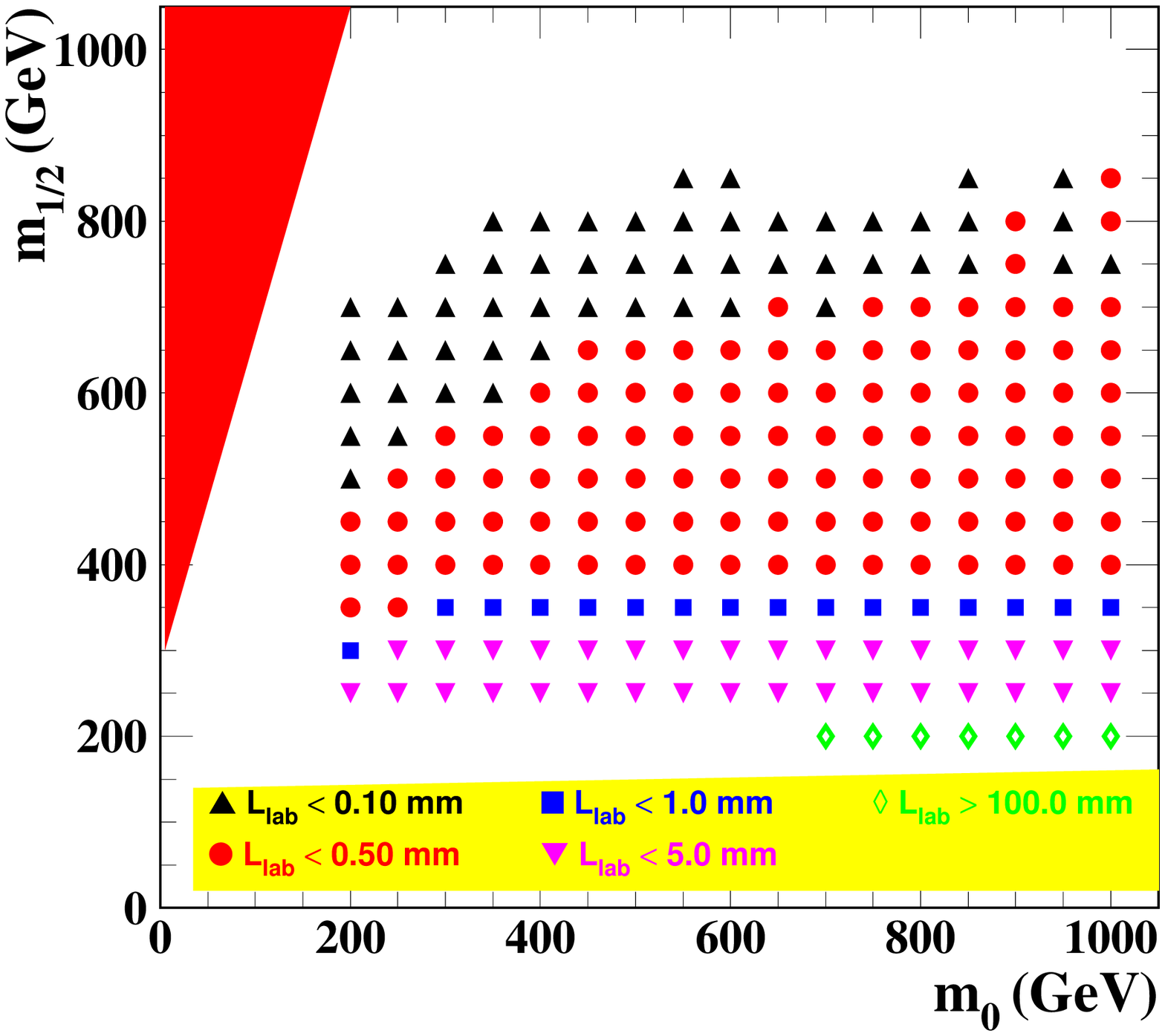}
\caption{Average distance traveled by the LSP in the laboratory frame
  as a function of the $m_0 \otimes m_{1/2}$ point for $A_0 = -100$
  GeV, $\tan \beta = 10$, and sgn $\mu > 0$.}
\label{fig:ldec_lab}
\end{figure}
%%%%%%%%%%%%%%%

In figure~\ref{fig:ldec_lab} we depict the average distance traveled
by the LSP as observed in the laboratory frame. As we can see, a
substantial fraction of the LSP decays takes place within the pixel
detector, except for very low $m_{1/2}$ values.  It is interesting to
notice that the pattern shown in the figure is similar to the one in
figure~\ref{fig:decaylength}, as we could easily expect. 
Since most of the LSP decays occur inside the beam pipe we can anticipate 
a small backgrounds associated to particles scattering in the 
detector material.
\smallskip

In order to obtain the LSP decay length ($L_0$) from the distance
traveled in the laboratory frame ($d$) we considered the $ m_{obs} d
/p_{obs}$ distribution, with $m_{obs}$ ($p_{obs}$) being the measured
invariant mass (momentum) associated to the displaced vertex, and then
we fitted it with an exponential
\[
    e^{-\frac{m_{obs} d}{p_{obs} L_0} }
\]
where the fitting parameter ($L_0$) is the LSP decay
length. \smallskip

\begin{figure}
\includegraphics[width=0.48\textwidth]{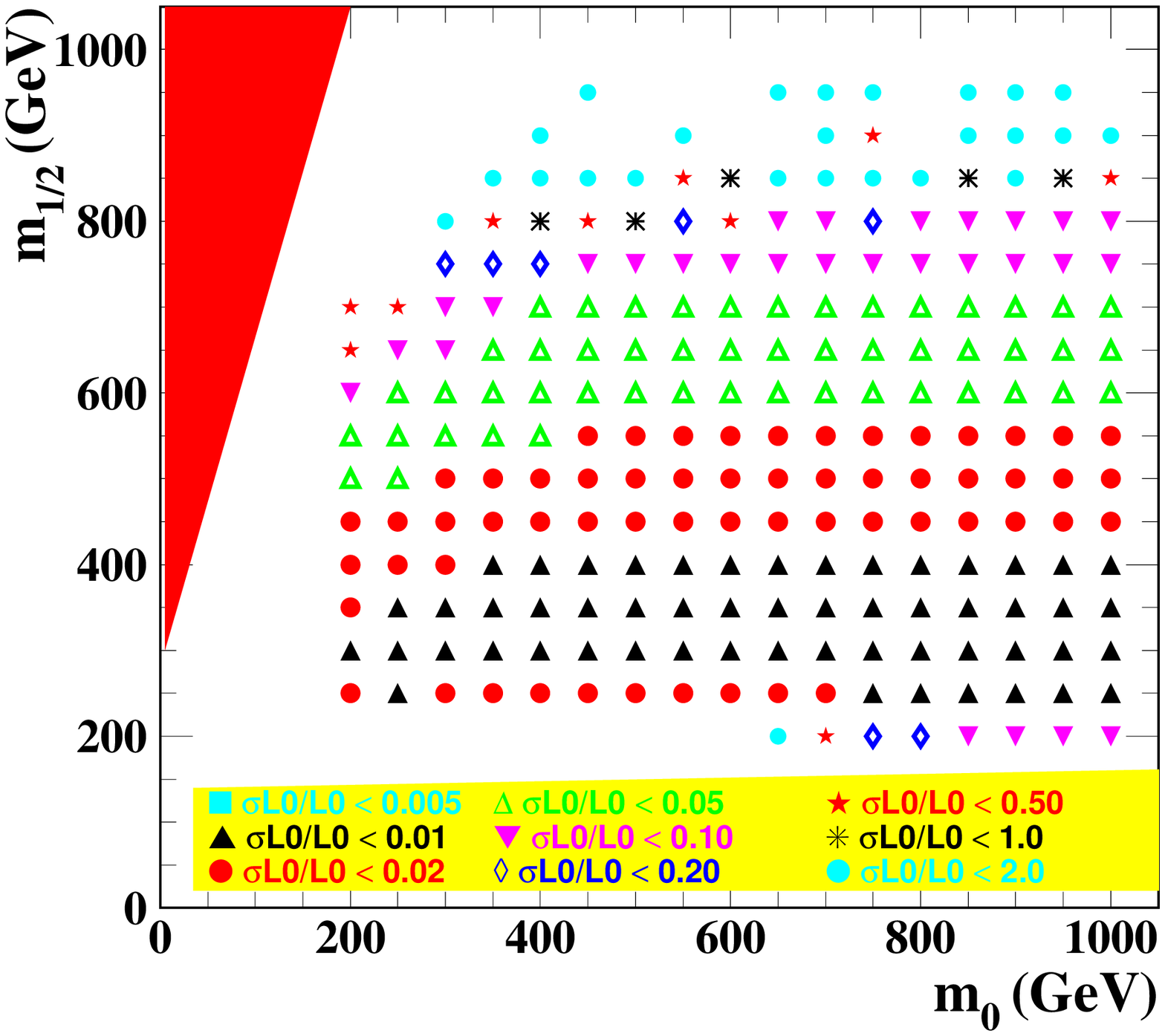}
\hfill
\includegraphics[width=0.48\textwidth]{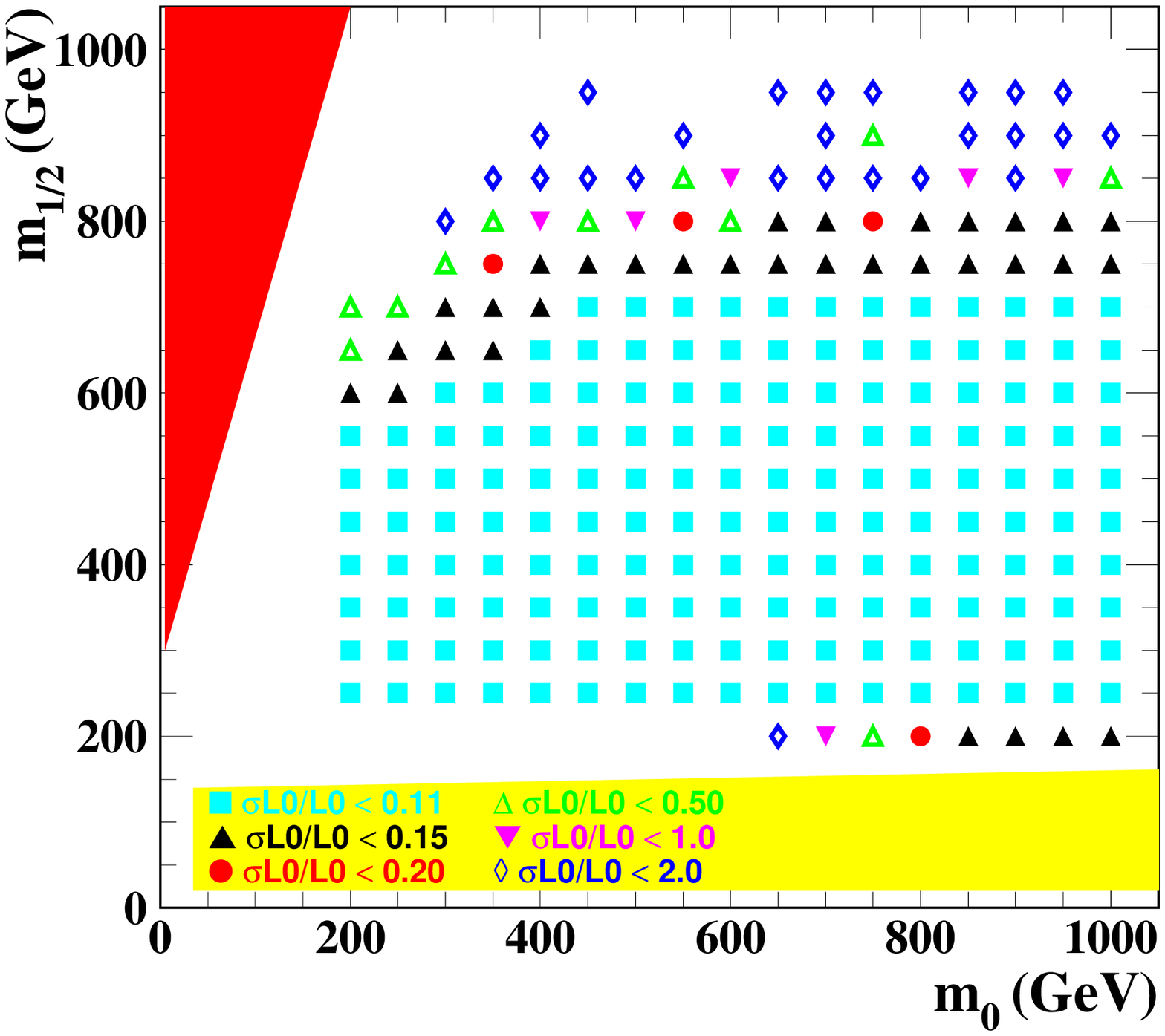}
\caption{ Relative error ($\sigma_{L_0}/L_0$) in the determination of
  the LSP decay length as a function of the $m_0 \otimes m_{1/2}$ for
  $A_0 = -100$ GeV, $\tan \beta = 10$, sgn $\mu > 0$, and an
  integrated luminosity of 100 fb$^{-1}$. The left (right) panel
  assumes no error (10\% error) in the measurement of distance
  traveled by the LSP.}
\label{fig:lsp_ldec}
\end{figure}

In order to disentangle the energy and momentum uncertainties and
  the statistical errors from the intrinsic limitation associated
to the tracking we first neglect the latter one.
In the left panel of figure~\ref{fig:lsp_ldec} we present the expected
precision in the decay length determination in the plane $m_0 \otimes
m_{1/2}$ for an assumed integrated luminosity of 100 fb$^{-1}$.
As one can see, these sources of error have a small impact in the
determination of the decay length, except for heavier LSP masses where
we run out of statistics. In fact, the contribution of these sources
of uncertainty is smaller than 5\% for neutralino masses up to 280 GeV
($m_{1/2} \simeq 700$ GeV).

Clearly, the actual achievable precision of LSP lifetime
determination at the LHC experiments depends on the
ability to measure the LSP traveled distance in the laboratory.
We present in the right panel of figure~\ref{fig:lsp_ldec} the
attainable precision on the decay length assuming a 10\% tracking
error \cite{ross:2011pr} in the LSP flight distance to get a rough
idea. Clearly the precision in the decay
length gets deteriorated, however, it is still better than 15\% within
a relatively large fraction of the parameter space under this assumption
but would get corresponding worse if this uncertainty were larger.

%%%%%%%%%%%%%%
%%%%%%%%%%%%%%%

%%%%%%%%%%%%%%%%%%%%%%%%%%%%%%%%%%%%%%%%%%%%%%%%%%%%%%%%%%%%%%%%%%%%%%
\section{LSP Branching ratio measurements}
\label{sec:lsp-braching-ratio}

As we have already mentioned, the neutrino mass squared difference
$\Delta m_{32}^2$ controls the ratio given in Eq.~\ref{eq:ratio},
therefore we should also study how well the neutralino LSP decay ratio
into $\ell W$ and $\nu Z$ can be determined.
In order to illustrate the LHC capabilities in probing LSP properties
at high energies we present the reconstruction efficiency for the
benchmark scenario
\[
  m_{1/2} = 250 \hbox{ GeV } \;\;\hbox{and} \;\;
 m_0 = 250 \hbox{ GeV}
\]
that yields a rather light LSP ($m_{\rm LSP}\simeq 101$ GeV) and heavy
scalars. For this point in parameter space the LSP possesses a decay
length $c\tau = 30$ $\mu$m and its dominant decay modes have the
following branching ratios:
\[
\begin{array}{lcrlcrlcr}
    \BR(\tilde\chi_1^0\to W^\pm e^\mp)&=&0.2\%, 
  & \BR(\tilde\chi_1^0\to W^\pm \mu^\mp)&=&27.6\%,
  & \BR(\tilde\chi_1^0\to W^\pm\tau^\mp)&=&31.3\%,
\\
    \BR(\tilde\chi_1^0\to b\bar{b}\nu)&=&7.1\%,
  & \BR(\tilde\chi_1^0\to Z \nu)&=&11.9\%,
  & 
\\
    \BR(\tilde\chi_1^0\to e^\pm\tau^\mp\nu)&=&5.5\%,
  & \BR(\tilde\chi_1^0\to\mu^\pm\tau^\mp\nu)&=&5.5\%, 
  & \BR(\tilde\chi_1^0\to\tau^\pm\tau^\mp\nu)&=&9.5\%
\end{array}
%  \label{eq:brsol}
\]
%

%%%%%%%%%%%%%%%
\begin{table}
 \begin{tabular}{|l|l|l|l|l|l|}\hline
    $N_{eqq'}$  &$N_{\mu qq'}$  & $N_{\tau qq'}$ &
    $N_{e\tau\nu}$  &$N_{\mu\tau\nu}$ &$N_{\tau\tau\nu}$
\\ 
\hline
 0.291  &   0.106   &  0.011  &  0.087    & 0.126  &  0.061
\\
\hline
 \end{tabular}
 \caption{Reconstruction efficiencies for neutralino LSP
   decays for our benchmark point. For the $\tau$ lepton
   only hadronic final states have been considered while the $\tau$ decays 
   into electrons and muons were included in the first two
entries. }
\label{tab1}
\end{table}
%%%%%%%%%%%%%%%

We present in Table~\ref{tab1} the reconstruction
efficiencies of the  LSP decay modes for our
chosen benchmark point. The reconstruction efficiencies for final
states containing $\tau$'s are much smaller, as expected, leading to a
loss of statistics in these final states. For an exhaustive study of
the reconstruction efficiencies see Ref.~\cite{DeCampos:2010yu}.
\smallskip

\begin{figure}[t]
  \centering
 \includegraphics[width=0.48\textwidth]{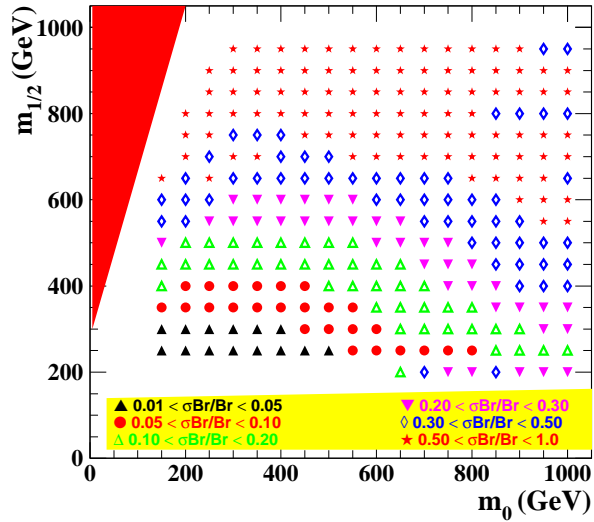}
 \caption{Expected error on the $\hbox{Br} ( \tilde{\chi}^0_ 1 \to
   \ell W + \nu Z) $ as a function $m_0 \otimes m_{1/2}$ for an
   integrated luminosity of 100 fb$^{-1}$. }
\label{fig:br}
\end{figure}

We present in figure~\ref{fig:br} the expected error on the LSP
branching ratio $\hbox{Br} ( \tilde{\chi}^0_ 1 \to \ell W + \nu Z) $
as a function $m_0 \otimes m_{1/2}$ for an integrated luminosity of
100 fb$^{-1}$. In order to evaluate this error we studied the
reconstruction efficiency for this final state and simulated 100
fb$^{-1}$ of data for several points in the $m_0 \otimes m_{1/2}$
plane. As one can see, this branching ratio can be well determined in
the regions of large production cross section, {\em i.e.} small $m_0$
and $m_{1/2}$. Although for heavier neutralinos the precision
diminishes, still this branching ratio can be determined to within
20\% in a large portion of the parameter space.
In order to study the possibility of LHC to probe the atmospheric mass, 
we have evaluated 
Br$(\tilde{\chi}^0_1 \rightarrow W\ell)$ + Br$(\tilde{\chi}^0_1 \to Z \nu)$ 
appearing in Eq.~\ref{eq:ratio}. 
The $W\ell$ channel is obtained by first reconstructing 
displaced vertices with hadronic $W$ decays, $jj\ell$, in the final state.
Beside the cuts described in sections \ref{ana:frame} and
\ref{sec:lsp-mass-measurement} we have
applied an invariant mass cut on the jet pair: $|M_{W} -M_{jj}| <20$ GeV
to disentangle the $W$-contribution to this final state.
Afterward we get the branching ratio for $W\ell$  using
\begin{equation}
\mathrm{Br}(\tilde{\chi}^0_1 \rightarrow W\ell) = 
\frac{\mathrm{Br}(\tilde{\chi}^0_1 \rightarrow jj\ell)}{N_{lqq'}}
\times\left(1+\frac{\mathrm{Br}(W \rightarrow \ell\nu)}
              {\mathrm{Br}(W \rightarrow qq\prime)} \right)\;\;.
\end{equation} 
The $Z\nu$ channel was calculated similarly by reconstructing 
the displaced vertices with hadronic $Z$ 
decays, $jj\nu$, in the final state and properly rescaling it.
Also here we have applied an invariant mass cut on the jet pair:
$|M_{Z}-M_{jj}| < 20$ GeV.

%%%%%%%%%%%%%%%
%%%%%%%%%%%%%%%

%%%%%%%%%%%%%%%%%%%%%%%%%%%%%%%%%%%%%%%%%%%%%%%%%%%%%%%%%%%%%%%%%%%%%%
\section{LSP properties and atmospheric neutrino oscillations}
\label{sec:lsp-prop-neutr}

\begin{figure}[h!]
   \centering
  \includegraphics[width=0.45\textwidth]{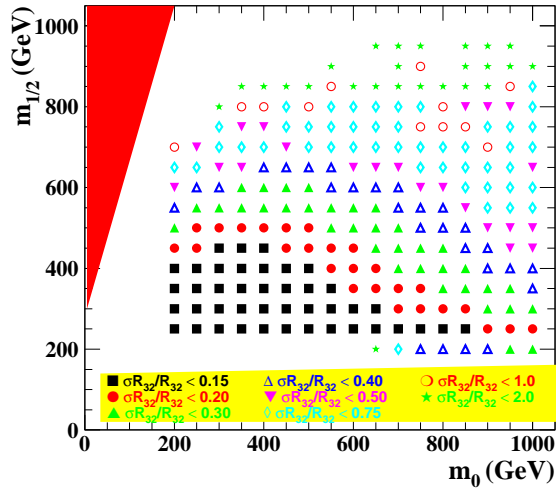}
  \caption{Expected accuracy on the ratio $R_{32}$ as a function of
    $m_0 \otimes m_{1/2}$ for an integrated luminosity of 100
    fb$^{-1}$.  }
 \label{fig:dm32}
\end{figure}

As seen in Section II, the MSSM augmented with bilinear
$R$--parity violation exhibits correlations between LSP decay
properties and the neutrino oscillation
parameters~\cite{Hirsch:2000ef,diaz:2003as}, which are by now well
measured in neutrino oscillation experiments~\cite{Tortola:2012te}.
In particular the squared mass difference $\Delta m_{32}^2$ is
connected to the ratio $R_{32}$ between the LSP decay length and its
branching ratio into $\ell W$ and $\nu Z$; see the right panel of
figure~\ref{fig:correlations}.
In figure~\ref{fig:dm32} we display the expected accuracy on the ratio
$R_{32}$ as a function of $m_0 \otimes m_{1/2}$ for an integrated
luminosity of 100 fb$^{-1}$ and assuming 10\% precision in the
determination of the LSP traveled distance.
As we can see $R_{32}$ can be determined with a
precision 20--30\% in a large fraction of the $m_0 \otimes m_{1/2}$
plane and, as expected, the precision is lost for heavy LSPs.  For
small LSP masses the error on $R_{32}$ is dominated by the uncertainty
on the decay length, while for heavier LSPs the dominant contribution
comes from the branching ratio determination due to the limited
statistics.  \smallskip

%%%%%%%%%%%%%%%
%%%%%%%%%%%%%%%

It is interesting to notice from the right panel of
figure~\ref{fig:correlations} that a measurement of $R_{32}$ with
20--30\% precision it is enough to determine the correct magnitude of
$\Delta m_{32}^2$ using the BRPV-mSUGRA framework. Nevertheless, a
much higher precision is needed to obtain uncertainties similar to the
neutrino experiments such as MINOS/T2K~\cite{Tortola:2012te}. On the
other hand, the relation between the atmospheric mixing angle and the
ratio of the LSP branching ratios into $\tau W$ and $\mu W$ can lead
to more stringent tests of the BRPV--mSUGRA model. In
Ref.~\cite{DeCampos:2010yu} it was shown that this ratio can be
determined at the LHC with a precision better than 20\% in a large
fraction of the $m_0 \otimes m_{1/2}$ plane. From
figure~\ref{fig:correlations} we can see that this precision is enough
to have a determination for $\tan^2 \theta_{23}$ with an error similar
to the low energy neutrino oscillation measurements. 
Looking from a different point of view, the collider data can be combined with
neutrino data to determine the underlying parameters of the model. In this case
collider and neutrino data give 'orthogonal' information as has been shown in
\cite{Thomas:2011kt}.
\smallskip

%%%%%%%%%%%%%%%%%%%%%%%%%%%%%%%%%%%%%%%%%%%%%%%%%%%%%%%%%%%%%%%%%%%%%%
\section{Conclusions}
\label{sec:conclusions}

We have analyzed the LHC potential to determine the LSP properties,
such as mass, lifetime and branching ratios, within minimal
supergravity with bilinear $R$-parity violation.  We saw that the LSP
mass determination is rather precise, while the LSP lifetime and
branching ratios can be determined with a 20\% error in a large
fraction of the parameter space. This is enough to allow for
qualitative test of the BRPV--mSUGRA model using the $R_{32}$--$\Delta
m_{32}^2$ correlation.
On the other hand, semi-leptonic LSP decays to muons and taus correlate
extremely well with neutrino oscillation measurements of
$\theta_{23}$. \smallskip

In the BRPV model for low values of $M_{1/2}$ one can have sizeable
branching ratios into the final states $e\tau\nu$ and
$\mu\tau\nu$. These decays are potentially interesting for testing
another aspects of the model associated with solar neutrino
physics. As shown in \cite{diaz:2003as} in regions of parameter space
where the scalar taus are not very heavy, usually the loop with
taus-staus in the diagram dominates the 1-loop neutrino mass. In this
case the solar angle is predicted to be proportional to
$(\tilde\epsilon_1/\tilde\epsilon_2)^2 \propto \tan^2\theta_{\odot}$.
Here, $\tilde\epsilon = V_{\nu}^{T,tree} {\vec \epsilon}$, with
$V_{\nu}^{T,tree}$ being the matrix which diagonalizes the tree-level
neutrino mass. Note that $V_{\nu}^{T,tree}$ is entirely determined in
terms of the $\Lambda_i$. In the BRPV model, RPV couplings of the
scalar tau are proportional to the superpotential parameters
$\epsilon_i$. Ratios of the decays Br$(\chi^0_1 \to
e\tau\nu)/$Br$(\chi^0_1 \to \mu\tau\nu)$ are then given, to a very
good approximation by Br$(\chi^0_1 \to e\tau\nu)/$Br$(\chi^0_1 \to
\mu\tau\nu) \propto (\epsilon_1/\epsilon_2)^2$. If the $\Lambda_i$
where known, this could be turned into a test of the prediction for
the solar angle.  Note that in the limit where the reactor angle is
exactly zero and the atmospheric angle exactly maximal one obtains
$(\tilde\epsilon_1/\tilde\epsilon_2)^2= 2(\epsilon_1/\epsilon_2)^2$.
However, the $\Lambda_i$ are currently not well fixed, due to the
comparatively large uncertainty in the atmospheric angle.  Thus the
correlation between three-body leptonic decays of the
neutralino with tau final states and the solar angle has a
rather large uncertainty. 
This prevents a stringent consistency test of the model using
 these decays.

All in all we have shown that neutralino decays can be used to
  extract some of their properties rather well in models with bilinear
  $R$-parity violation. Properties such as the decay length and the
  ratio of semi-leptonic decay branching ratios to muons and taus
  correlate rather well with atmospheric neutrino oscillation
  parameters. These features should also apply to schemes where the
  gravitino is the LSP and the neutralino is the next to lightest SUSY
  particle~\cite{Hirsch:2005ag,Restrepo:2011rj}. For gravitino masses in the allowed
  range where it plays the role of cold dark matter, its R-parity
  conserving decays are negligible compared to its R parity violating
  decays. The latter follow the same patter studied in the present
  paper, so that the results derived here should also hold.

%%%%%%%%%%%%%%%%%%%%%%%%%%%%%%%%%%%%%%%%%%%%%%%%%%%%%%%%%%%%%%%%%%%%%%
\section*{Acknowledgments}

We thank Susana Cabrera, Vasiliki Mitsou and Andreas Redelbach
for useful discussions on
the ATLAS experiment. 
W.P.~thanks the IFIC for hospitality during an extended stay.
Work supported by the Spanish MINECO under grants FPA2011-22975 and
MULTIDARK CSD2009-00064 (Consolider-Ingenio 2010 Programme), by
Prometeo/2009/091 (Generalitat Valenciana), by the EU ITN UNILHC
PITN-GA-2009-237920. O.J.P.E is supported in part by Conselho Nacional
de Desenvolvimento Cient\'{\i}fico e Tecnol\'ogico (CNPq), by
Funda\c{c}\~ao de Amparo \`a Pesquisa do Estado de S\~ao Paulo
(FAPESP) and in part by the European Union FP7 ITN INVISIBLES (Marie
Curie Actions, PITN- GA-2011- 289442) network.  W.P.\ has been
supported by the Alexander von Humboldt foundation and in part by the
DFG, project no.\ PO-1337/2-1.

%%%%%%%%%%%%%%%%%%%%%%%%%%%%%%%%%%%%%%%%%%%%%%%%%%%%%%%%%%%%%%%%%%%%%%

% \bibliographystyle{h-physrev5}
% \bibliography{merged}

%%%%%%%%%%%%%%%%%%%%%%%%%%%%%%%%%%%%%%%%%%%%%%%%%%%%%%%%%%%%%%%%%%%%%%
\end{document}